\documentclass[twocolumn,showpacs,preprintnumbers,amsmath,amssymb,pra,showkeys]{revtex4}

\usepackage{txfonts}
\usepackage{amsmath}

\usepackage{graphicx}
\usepackage{dcolumn}
\usepackage{bm}

\begin{document}


\title{Dynamical Casmir effect and Conductivity}

\author{Xiao-Min Bei}
\author{Zhong-Zhu Liu}%
 \email{xiaominbei@gmail.com}
\affiliation{%
Department of Physics, Huazhong University of Science and Technology,
Wuhan, 430074, China
}%

\date{\today}

\begin{abstract}
In this paper we find that the second law of thermodynamics requires an upper limit of the conductivity. To begin with we present an ideal model, the cavity with a mobile plate, for studying the thermodynamic properties of radiation field. It is shown that the pressure fluctuation of thermal radiation field in the cavity leads to the random motion of the plate and photons would be generated by dynamical Casimir effect. Meanwhile, such photons obey a non-thermal distribution. Then, to ensure the second law of thermodynamics, there must be a upper limit of the conductivity.
\end{abstract}

\pacs{?}

\maketitle

\section{Introduction}

When the space symmetry of a system is broken the statistical fluctuation may lead to macroscopic motion, which is called molecular motor \cite{1,2,3,4,5}. At the same time, the system must obey the second law of thermodynamics. This will give the system of molecular motor some rigorous limit. Contrasting to the space symmetry, it has recently paid more attention to the time asymmetry of the system. People want to understand  what macroscopical phenomena will result from thermal fluctuations and what constraints on the system will be prescribed by the second law of thermodynamics.

In quantum field theory, time-dependent boundary conditions or uncharged mirrors in accelerated motion may induce particle creation, even when the initial state of a quantum field is the vacuum \cite{6,7,8}. This purely quantum effect has been known as the dynamical Casimir effect (DCE) \cite{9}. In this case, along with moving boundaries, the vacuum state of the electromagnetic field is changed, and the changing vacuum results in the generation of photons. However, the reverse process does not occur. This is the time symmetry breaking of the system, which will lead to a phenomenon similar to the molecular motor.

Generated photons in the DCE obey a non-thermal distribution. In some peculiar case, the non-thermal distribution will lead to a system's macroscopically breaking of the thermal equilibrium. To keep the thermal equilibrium of the system, the second law of thermodynamics requires a compensatory effect. In this work, we shall study how need material to absorb photons generated by the DCE for ensuring the second law of thermodynamics. Firstly we present an ideal model, a conductor cavity with a mobile plate inside it. Assuming the system to be initially at thermal equilibrium, the pressure fluctuation will result in a pressure difference on both sides of plate. It leads to a random motion of the conducting plate and photons would be generated in the cavity by the DCE. Note that these photons created do not satisfy the thermal distribution but the super-Poissonian distribution \cite{10}, which may break the balance of thermal radiation field. To ensure the second law of thermodynamics, created photons should be absorbed by the cavity wall in the relaxaion time; that is to say, the photon absorption rate has to be greater than that of the generation rate. Meanwhile, both the photon absorption rate and the generation rate are related with the matieral's conductivity. Thus an upper limit of the conductivity is prescribed.

It is confirmed experientially already that there is an upper limit of the conductivity. However, this existence of the upper limit lacks a theoretical proof. In this paper, we shall prove this upper limit according to the second law of thermodynamics.

The article is organized as follows. In Sec. II, we start by analyzing the pressure fluctuation of the thermal radiation in the cavity. Then we establish a Langevin equation describing the plate motion and derive the time correlation function of the acceleration. In Sec. III, according to DCE we obtain the numbers of photons created by random motion of the conducting plate, and present an expression of the relative photon generation rate per volume. Next, the second law of thermodynamics is considered in Sec. IV, and we show that requiring the photon absorption rate to be greater than the generation rate, it could lead to an upper limit of the conductivity. Finally we conclude our work in the last section.

\section{The Model}

In this section we will present a thermodynamic analysis of thermal radiation field in the cavity with a mobile plate inside it.

We consider a three-dimensional model of a thermal radiation field within a rectangular cavity with conducting walls. The cavity has the dimensions $L_x$, $L_y$ and $L_z$. At the midpoint of the cavity $\left(x=L_x /2\right)$ a thin mobile plate is located, which is made of the same material as the cavity walls. Thus the cavity is divided into two regions: region I $\left(0 \le x\le {L_x /2} \right)$ and region II $\left({L_x/2}\le x \le L_x \right)$. For the sake of simplicity we assume the system to be initially at thermal equilibrium corresponding to some nonvanishing temperature $T$ and the temperature and pressure within two regions are equal, i.e., $T_\texttt{I}=T_{\texttt{II}}$, and $p_\texttt{I}=p_{\texttt{II}}$. Owing to the pressure fluctuation in respect of the thermal equilibrium, a pressure difference on both sides of plate emerges \cite{10}. The pressure difference is variable stochastically and leads to the random motion of the conducting plate. Under the motion of the plate, photons would be generated in the cavity by the DCE.

Initially, the pressure of back-body radiation in the cavity can be simply expressed by \cite{11}
\begin{equation}
\label{eq1}
p_a=\frac{4}{3}\sigma T^4,
\end{equation}
where the coefficient $\sigma$ is the Stefan-Boltzmann constant and the subscript $a=\texttt{I},\texttt{II}$ differentiate the radiation pressures in the left or right cavity. Note that this pressure is proportional to the fourth power of the temperature. At the same time, owing to the pressure fluctuation, we can take the following expression for the mean square fluctuation of the pressure \cite{11}
\begin{equation}
\label{eq2}
\overline{\left(\Delta p_a\right)^2}=-k_BT\frac{\partial p_a}{\partial V}\bigg|_S,
\end{equation}
where $k_B$ is Boltzmann's constant, and $V$ is the cavity volume.

Now we put Eq. (\ref{eq1}) into the above formula (\ref{eq2}) and take into account of Maxwell relations \cite{11}. Then the mean square fluctuation of this pressure can be rewritten as
\begin{equation}
\label{eq3}
\overline{\left(\Delta p_a\right)^2}=\frac{\alpha}{3V}k_BT^5,
\end{equation}
with $\alpha=\frac{4\sigma}{c}$. It is shown that the mean square fluctuation of the pressure is proportional to the fifth power of the temperature and inversely proportional to the volume.

There is a correlation between $\Delta p_a\left(t\right)$ at different instants. This means that the value of $\Delta p_a$ at a given instant $t$ affects the probabilities of its various values at a later instant $t'$. We can characterize the time correlation by the mean value of the product \cite{11}
\begin{equation}
\label{eq4}
\overline{\Delta p_a(t)\Delta p_a(t')}=\overline{[\Delta p_a]^2}e^{-\lambda(t'-t)}.
\end{equation}
Here the constant $1/\lambda$ determines the order of magnitude of the relaxation time for the establishment of complete equilibrium.

Since pressure fluctuations on both sides of the plate are obviously irrelevant of each other, the fluctuations of the pressure $\Delta p_\texttt{I}(t)$ and $\Delta p_{\texttt{II}}(t)$ are statistically independent at each instant. Then the pressure difference on both sides of plate can be simply written as $\Delta P(t)=\Delta p_\texttt{I}(t)-\Delta p_{\texttt{II}}(t)$ which is also a random variable. Finally, using Eq. (\ref{eq4}), we obtain a formula for the time correlation of pressure difference:
\begin{equation}
\label{eq5}
\overline{\Delta P(t)\Delta P(t')}=\frac{2\alpha}{3V}k_BT^5e^{-\lambda(t'-t)},
\end{equation}
At the same time, in the formal limit $t'\rightarrow t$, the mean square fluctuation of the pressure difference is given simply as
\begin{equation}
\label{eq6}
\overline{[\Delta P]^2}=\frac{2\alpha}{3V}k_BT^5.
\end{equation}
To illustrate how we may incorporate dynamics into the discussion of a stochastic process, let $\dot{x}(t)$ be the velocity of the plate. The Newtonian equation of the plate's motion is given by
\begin{eqnarray}
\label{eq7}
M\ddot{x}+\beta\dot{x}=\Delta P(t)S+\frac{1}{2}\left(\frac{\partial A\left(x,t\right)}{\partial x}\right)\left(\big|_{x=L_0/2}-\big|_{x=L_0/2+\Delta x}\right),
\end{eqnarray}
where $M$ and $S$ are the mass and area of the plate respectively, $\beta$ is the friction coefficient, and $\Delta x$ is the thickness of the plate. This is called the Langevin equation \cite{12,13}. The first term $\Delta P(t)S$ denotes a randomly fluctuating external force induced by the pressure difference while the second term, which is enough small to can be neglected \cite{14}, represents the recoil force due to photon radiation. For simplicity, the above formulas can be equivalently expressed as
\begin{equation}
\label{eq8}
\ddot{x}+\gamma\dot{x}=s\Delta P(t),
\end{equation}
where $s=\frac{S}{M}$ and $\gamma=\frac{\beta}{M}$ is a damping coefficient. The above equation (\ref{eq8}) is a random differential equation describing how the plate takes the Brownian movement. Let us assume that at time $t=0$, the velocity and position of the plate are $\dot{x}(0)$ and $x(0)$, respectively. Then the solution of the Eq. (\ref{eq8}) is
\begin{equation}
\label{eq9}
\dot{x}(t)=\dot{x}(0)e^{-\gamma t}+e^{-\gamma t}\int_0^ts\Delta P\left(\xi\right)e^{\gamma\xi}d\xi,
\end{equation}
This equation gives $\dot{x}(t)$ for a single realization of $\Delta P\left(t\right)$. Since $\Delta P\left(t\right)$ is a stochastic variable, $\dot{x}(t)$ and $\ddot{x}(t)$ are also stochastic whose properties are determined by $\Delta P\left(t\right)$. The average velocity is $\overline{\dot{x}(t)}=\dot{x}(0)e^{-\gamma t}$. Using Eq.(\ref{eq5}) and Eq. (\ref{eq9}) , the time correlation function of the velocity can be obtained as follows
\begin{eqnarray}
\label{eq10}
\overline{\dot{x}(t)\dot{x}(t')}=&&\left(\dot{x}(0)\right)^2e^{-\gamma(t+t')}\nonumber\\
&&+s^2\overline{[\Delta P]^2}\left(\frac{e^{\lambda t}-e^{-\gamma t}}{\gamma+\lambda}\right)\left(\frac{e^{-\lambda t'}-e^{-\gamma t'}}{\gamma-\lambda}\right).
\end{eqnarray}
Then the time correlation function of the acceleration can be written in terms of Eq.(\ref{eq8})
\begin{equation}
\label{eq11}
\overline{\ddot{x}(t)\ddot{x}(t')}=\gamma^2\overline{\dot{x}(t)\dot{x}(t')}+s^2\overline{\Delta P(t)\Delta P(t')},
\end{equation}
where the formula $\overline{\Delta P}=0$ is used. Now we put Eq. (\ref{eq5}) and Eq. (\ref{eq10}) into the above equation (\ref{eq11}), and then derive the time correlation function of the acceleration as
\begin{eqnarray}
\label{eq12}
\overline{\ddot{x}(t)\ddot{x}(t')}=&&\gamma^2\left(\dot{x}(0)\right)^2e^{-\gamma(t+t')}\nonumber\\
&&+\gamma^2s^2\overline{[\Delta P]^2}\left(\frac{e^{\lambda t}-e^{-\gamma t}}{\gamma+\lambda}\right)\left(\frac{e^{-\lambda t'}-e^{-\gamma t'}}{\gamma-\lambda}\right)\nonumber\\
&&+s^2\overline{[\Delta P]^2}e^{-\lambda(t'-t)}.
\end{eqnarray}
For time long enough, namely $t,t'\gg1/\gamma$, the initial velocity of the plate can be neglected. Then Eq. (\ref{eq12}) becomes simply,
\begin{eqnarray}
\label{eq13}
&&\overline{\ddot{x}(t)\ddot{x}(t')}\nonumber\\
&&=s^2\overline{[\Delta P]^2}\left(\gamma^2\frac{e^{-\lambda(t'-t)}-e^{\lambda t-\gamma t'}-e^{-\gamma t-\lambda t'}+e^{-\gamma(t+t')}}{\gamma^2-\lambda^2}+e^{-\lambda(t'-t)}\right).\nonumber\\
\end{eqnarray}
In the following we can take $\ddot{x}(t)$ to be any random function of $t$, and shall write $t$ instead of $t'$ for convenience. Then one has
\begin{eqnarray}
\label{eq14}
\overline{[\ddot{x}(t)]^2}=s^2\overline{[\Delta P]^2}\left(\gamma^2\frac{1-e^{(\lambda-\gamma)t}-e^{-(\lambda+\gamma)t}+e^{-2\gamma t}}{\gamma^2-\lambda^2}+1\right).
\end{eqnarray}
Thus it is equal to the mean square value of the fluctuating function $\ddot{x}(t)$.

Hence, the radiation pressure fluctuations in two different regions of the cavity would lead to the random motion of the conducting plate. In contrast to \cite{10}, we are interested here in whether it is likely to create photons. This will be discussed in detail in the next section.

\section{Random Motion And Photon Creation}

In above section, we get the time correlations of the acceleration and the velocity of the plate. The movement of the plate brings on the photon generation according to the theory of the DCE \cite{6,7}. And the relative photon generation rate per volume will be derived. First of all we will calculate the numbers of photons created by the DCE in the right or left cavity, which is due to random motion of the conducting plate. Comparing with Ref. \cite{10}, we have assumed that the type of motion of the plate may be described by
\begin{eqnarray}
\label{eq15}
x(t)=\left\{
       \begin{array}{ll}
         x_1(t)=L & \hbox{$t\leq 0$} \\
         x_2(t) & \hbox{$0\leq t \leq t_1$} \\
         x_3(t)=x_2(t)=x_0 & \hbox{$t>t_1$}
       \end{array}
     \right.,
\end{eqnarray}
with $L=\frac{L_x}{2}$, and $x_0$ is a constant satisfied $0<x_0<L_x$. Note that the cavity is at rest for times $t\leq 0$ and $t>t_1$, and we require $x_2(0)=L$, $\dot{x}_2(0)=0$, $\dot{x}_3(t_1)=0$, and $x(t)\geq0$ for all $t$Since the system is initially in thermal equilibrium at temperature $T$, we may take a number state $|n_k\rangle$ to be the initial quantum state of photons
\begin{equation}
\label{eq16}
|n_k\rangle=\frac{\left(a_k^{(1)\dag}\right)^{n_k}}{\sqrt{n_k!}}|0\rangle,
\end{equation}
where it denotes a number state with $n_k$ quanta in the $k$th mode and $a_k^{(1)\dag}$ are the creation operators at $t\leq 0$. Then the mean number of photons created from vacuum in this state during the time interval $[0,t_1]$ is \cite{10}
\begin{eqnarray}
\label{eq17}
&&\langle n_k|a_n^{(3)\dag}a_n^{(3)}|n_k\rangle\approx n_k\left\{\delta_{nk}\right.\nonumber\\
&&\ \left.+\frac{36n^2}{c^4\pi^4}\frac{1-\delta_{nk}}{(n-k)^6}
\left[\sqrt{\frac{n}{k}}\frac{1}{6}x_0\ddot{x}(t_1)-\sqrt{\frac{k}{n}}\frac{1}{6}x_0\ddot{x}(0)\right]^2\right\}.
\end{eqnarray}
where $a_n^{(3)}$ and $a_n^{(3)\dag}$ are the annihilation and creation operators at $t>t_1$. Substituting Eqs. (\ref{eq13}) and (\ref{eq14}) into the expansion (\ref{eq17}) we can obtain the ensemble average of the expectation value of the number operator
\begin{equation}
\label{eq18}
\overline{\langle n_k|a_n^{(3)\dag}a_n^{(3)}|n_k\rangle}\approx n_k\left\{\delta_{nk}+\frac{1-\delta_{nk}}{(n-k)^6}A(n,k)\right\},
\end{equation}
where
\begin{eqnarray}
\label{eq18'}
&&A(n,k)=\frac{n^2}{c^4\pi^4}x_0^2s^2\frac{2k_B\alpha T^5}{3V}\left\{\left(\frac{n}{k}+\frac{k}{n}-2e^{-\lambda t_1}\right)\right.\nonumber\\
&&\ \ \left.+\frac{\gamma^2}{\gamma^2-\lambda^2}\left[\frac{n}{k}\left(1-e^{(\lambda-\gamma)t_1}+e^{-2\gamma t_1}-e^{-(\gamma+\lambda)t_1}\right)\right]\right\}.
\end{eqnarray}
Note that from Eq. (\ref{eq18}) the statistics of photons created do not satisfy a thermal distribution but a super-Poissonian distribution \cite{10}. Therefore, it turns out that photons generated in the cavity are different from thermal photons. In this case the relative photon creation rate per volume in the $k$th mode can be written in the form
\begin{eqnarray}
\label{eq19}
\tilde{P}_k(t_1)&&=\sum_n\frac{1}{t_1}\frac{1}{n_k}\left[\overline{\langle n_k|a_n^{(3)\dag}a_n^{(3)}|n_k\rangle}-n_k\right]\nonumber\\
&&\approx\sum_n\frac{1-\delta_{nk}}{t_1(n-k)^6}A(n,k).
\end{eqnarray}
With the proviso that $t_1\gg1/\gamma$ and $t_1\gg1/\lambda$ the total photon creation rate per volume in the lowest mode with $k=1$ and $n=2$ are
\begin{eqnarray}
\label{eq20}
\tilde{P}(t_1)&&=\sum_k\tilde{P}_k(t_1)\nonumber\\
&&\approx\frac{2x_0^2s^2\overline{[\Delta P]^2}}{t_1c^4\pi^4}\left[\frac{\gamma^2}{\lambda^2-\gamma^2}\left(e^{(\lambda-\gamma)t_1}-1\right)+\frac{5}{4}\right].
\end{eqnarray}
We note that this equation is greater than zero in any case. Therefore, over time the total number of photons created would continue to be generated and increased by the DCE. To keep the thermal equilibrium of the system, it is necessary to find a compensatory effect. This we shall discuss more fully in Sec.¢ô.

\section{Upper Limit Of The Conductivity}

From the previous discussion, we learn that thermal photons in the cavity satisfy the Planck distribution, and photons created by the DCE satisfy the super-Poissonian distribution. Thus, these two types of photons have different distributions and are statistically independent each other. From Eq. (\ref{eq20}) the total number of photons generated would continue to be increased as time goes on. If there are no other physical effects of compensation, the system will transition to a non-equilibrium state. This will violate the second law of thermodynamics.

Photons created by the DCE can not be directly converted into thermal photons. Therefore, to meet the second law of thermodynamics, photons created must be absorbed by conducting walls or plate. That is to say, the reflection coefficient of the conducting walls and plate need be less than 1.

First the reflection coefficient of the conducting walls and plate is expressed as $R$. Then the absorption rate per unit time can be written as $1-R^f$ \cite{15}, where $f=c/l$ is folding times of a light beam in the cavity. At the same time, in order to satisfy the second law of thermodynamics, the photon absorption rate per unit time must be greater than that of the generation rate
\begin{equation}
\label{eq21}
\left(1-R^f\right)\geq \tilde{P}(t).
\end{equation}
And according to Ref. \cite{16}, the reflection coefficient can be written as $R\approx1-2\sqrt{\frac{2\omega\varepsilon_0}{\sigma}}$. Inserting Eq.(\ref{eq20}) into Eq.(\ref{eq21}), we can get the expression of the conductivity as
\begin{eqnarray}
\label{eq22}
\sigma_c\leq\left(\frac{\sqrt{8\omega\varepsilon_0}ft_1c^4\pi^4}{2x_0^2s^2\overline{[\Delta P]^2}\left[\frac{\gamma^2}{\lambda^2-\gamma^2}\left(e^{(\lambda-\gamma)t_1}-1\right)+\frac{5}{4}\right]}\right)^2.
\end{eqnarray}
Note that the above expression (\ref{eq22}) represents an upper limit for the conductivity. For simplicity, different cases will be discussed. We shall complete our study by considering various limiting cases. If $\lambda\gg\gamma$, the relax time of radiation fluctuation $\lambda^{-1}$ is much less than the characteristic time $\gamma^{-1}$ of the plate motion. Then we can get
\begin{eqnarray}
\label{eq23}
\sigma_c\leq\left(\frac{\sqrt{8\omega\varepsilon_0}ft_1c^4\pi^4}{2x_0^2s^2\overline{[\Delta P]^2}\left(\frac{\gamma^2}{\lambda^2}e^{\lambda t_1}+\frac{5}{4}\right)}\right)^2.
\end{eqnarray}
In the opposite limiting case $\lambda\ll\gamma$, the results are
\begin{eqnarray}
\label{eq24}
\sigma_c\leq\left(\frac{\sqrt{8\omega\varepsilon_0}ft_1c^4\pi^4}{2x_0^2s^2\overline{[\Delta P]^2}}\frac{4}{9}\right)^2.
\end{eqnarray}
By this means we can get an upper limit of the conductivity which is result from the DCE. This result has no concern with the structure of conductor and the property of its material. In nature, this is an inevitable consequence that is caused by the compatibility between the second law of thermodynamics and the dynamical Casimir Effect. Then one can simply estimate the value of this upper limit.

Now we insert some explicit numbers to get the upper limits of the conductivity. The parameters are given as follow: the mass of the plate $M\sim0.01kg$, $\gamma=10^{-3}s^{-1}$ and the frequency of the lowest mode is $\omega_1\sim1GHz$ for $L\sim0.1m$. Then in the former case approximately $\sigma_c\leq10^{22}S\cdot m^{-1}$ would be obtained at room temperature $T\sim290K$ during 1 hour, where we take $\lambda=\frac{3(1-R)c}{2L}$ (as in the appendix). And  in the latter case $\sigma_c\leq10^{136}S\cdot m^{-1}$ would be gotten.

\section{Conclusion}

This paper discuss the possibility of upper limits of the conductivity by virtue of the compatibility between the second law of thermodynamics and the DCE. We first consider a three-dimensional model of a thermal radiation field within a rectangular cavity with a thin conducting plate. The system to be initially at thermal equilibrium, the pressure fluctuation in thermal equilibrium will result in a pressure difference on both sides of plate. Then we establish a Langevin equation describing the plate movement and derive the time correlation function of the acceleration. It is important to notice that the random motion of the conducting plate is in general nonzero, which may lead to photon generation even from vacuum.

The relative photon generation rate per volume is derived in Sec.III. First of all we calculate the numbers of photons created by the DCE in the right or left cavity, which is due to random motion of the conducting plate. And it is found that the statistics of photons created by DCE do not satisfy a thermal distribution but a super-Poissonian distribution. Thus, it turns out that Photons generated in this cavity obey a non-thermal distribution and present an expression of the relative photon generation rate per volume. Finally, to ensure the second law of thermodynamics, the photon absorption rate has to be greater than that of the generation rate. Thus we obtain this upper limit of the conductivity.

\section*{Appendix}

In this Appendix, we derive the expresssion of the coefficient $\lambda$ in the rectangular cavity with a thin conducting plate.

The energy density in the thermal equilibrium can be expressed as \cite{11}
\begin{eqnarray}
\label{eq25}
u=\frac{E}{V}=\alpha T^4.
\end{eqnarray}
And the intensity of emission is obtained as
\begin{eqnarray}
\label{eq26}
J=\frac{1}{4}c\alpha T^4.
\end{eqnarray}
We study in this paper how the thermal radiation field with a pressure fluctuation tends to equilibrium. The field only can exchange energy with the cavity wall in relaxation  time. The energy flux of the exchange energy is expressed by the temperature difference between the cavity wall and the thermal radiation field. So when the system comes back to equilibrium, the change in the energy density is given by
\begin{eqnarray}
\label{eq27}
\Delta u=\alpha T^4-\alpha T_0^4,
\end{eqnarray}
where $T$ and $T_0$ are the temperature of radiation field and cavity wall respectively. In this case the change in the intensity of absorption can be read from Eq. (\ref{eq26}) and Eq. (\ref{eq27})
\begin{eqnarray}
\label{eq28}
A\Delta J=\frac{1}{4}Ac\alpha T^4-\frac{1}{4}Ac\alpha T_0^4=\frac{1}{4}Ac\Delta u,
\end{eqnarray}
where $A$ is the absorptivity of the cavity wall. Then, for the rectangular cavity, the variation of internal energy per unit time can be represented as
\begin{eqnarray}
\label{eq29}
\frac{\Delta E}{\Delta t}=\tilde{S}A\Delta J,
\end{eqnarray}
where $\tilde{S}$ is the total surface area of the cavity. Thus we get the relaxation time for the establishment of complete equilibrium
\begin{eqnarray}
\label{eq30}
\Delta t=\frac{\Delta E}{\tilde{S}A\Delta J}.
\end{eqnarray}
Considering the coefficient $\lambda^{-1}$ being the order of magnitude of the relaxation time \cite{11}, we put Eq. (\ref{eq28}) into the above formula (\ref{eq30}) and then obtain
\begin{eqnarray}
\label{eq31}
\lambda=\frac{3Ac}{2L_x}=\frac{3(1-R)c}{2L_x}.
\end{eqnarray}
Here $R$ is the reflection coefficient and the equation $A=1-R$ has been used.

\begin{acknowledgments}
This work is supported in part by the National Natural Science Foundation of China (Grant No. 2010CB832800).
\end{acknowledgments}


\begin{thebibliography}{99}
\bibitem{1}	F. J\"{u}licher, A. Ajdari, and J. Prost,  Rev. Mod. Phys. {\bf69}, 1269 (1997).

\bibitem{2}	S. Muhuri and I. Pagonabarraga, Phys. Rev. E {\bf82}, 021925 (2010).

\bibitem{3} P. Reimann, Phys. Rep. {\bf361}, 57 (2002).

\bibitem{4} K. L. Sebastian, Phys. Rev. E {\bf61}, 937 (2000).

\bibitem{5} P. H\"{a}nggi and F. Marchesoni, Rev. Mod. Phys. {\bf81}, 387  (2009).


\bibitem{6} G. T. Moore, J. Math. Phys. {\bf11}, 2679 (1970); S. A. Fulling and P. C. W. Davies, Proc. R. Soc. London Ser. A {\bf 348}, 393 (1976); P. C. W. Davies and S. A. Fulling, Proc. R. Soc. London Ser. A {\bf356}, 237 (1977).

\bibitem{7}	V. V. Dodonov and A. B. Klimov, Phys. Rev. A {\bf53}, 2664 (1996).

\bibitem{8} A. Agnesi, C. Braggio, G. Bressi, G. Carugno, G. Galeazzi, F. Pirzio, G. Reali, G. Ruoso, and D. Zanello, J. Phys. A {\bf41}, 164024 (2008); A. Agnesi, C. Braggio, G. Bressi, G. Carugno, F. Della Valle, G. Galeazzi, G. Messineo, F. Pirzio, G. Reali,G. Ruoso, D. Scarpa, and D. Zanello, J. Phys.: Conf. Ser. {\bf161}, 012028 (2009).

\bibitem{9} C. K. Law, Phys. Rev. A {\bf49}, 433 (1994).

\bibitem{10} S. Sarkar, Quantum Opt. {\bf4}, 345 (1992).

\bibitem{11} L. D. Landau and E. M. Lifshitz, {\it Statistical Physics, Part 1} (Pergamon, New York, 1968).

\bibitem{12} J. H. Weiner, {\it Statistical Mechanics of Elasticity} (Dover, New York, 2002).

\bibitem{13} F. Reif, {\it Fundamentals of Statistical and Thermal Physics} (McGraw-Hill, New York, 1965).

\bibitem{14} C. K. Law, Phys. Rev. A {\bf51}, 2537 (1995).

\bibitem{15} S. C. Wu, Z. Z. Wan, H. Li, and Z.-Z. Liu, Chin. Phys. Lett. {\bf23}, 3173 (2006)


\bibitem{16} J. D. Jackson, {\it Classical Electrodynamics} (Wiley, New York, 1975).




\end{thebibliography}

\end{document}